# Ultralow thermal conductance across the [FePt/h-BN/FePt] interface

Chengchao Xu[1,2], Enbo Zhang[1,2], Bo-Yuan Yang[1,2], B.S.D.Ch.S. Varaprasad[1,2], David E. Laughlin[1,2,3], and Jian-Gang (Jimmy) Zhu[1,2,3,a)]

1) Data Storage Systems Center, Carnegie Mellon University, Pittsburgh, PA 15213, USA

2) Department of Electrical and Computer Engineering, Carnegie Mellon University, Pittsburgh, PA 15213, USA

3) Department of Materials Science and Engineering, Carnegie Mellon University, Pittsburgh, PA 15213, USA

a) Correspondence should be addressed to J. G. Zhu: jzhu@ece.cmu.edu

**ABSTRACT**

Heat transfer in nanocomposite materials has attracted great interest for various applications. Multilayer structures provide an important platform to study interfacial thermal transport and to engineer materials with ultralow thermal conductivity. Here we report on the fabrication and thermal characterization of [h-BN/L1$_0$-FePt]×N multilayers, where hexagonal boron nitride (h-BN) nanosheets (2.5 nm thick) and L1$_0$-FePt layers (6.5 nm thick) alternate periodically. Differential three-omega(3ω) measurements reveal an ultralow effective thermal conductivity of 0.60 ± 0.05 W·m$^{-1}$K$^{-1}$ across the multilayer films, and a low thermal boundary conductance (TBC) of 67.9 ± 6.6 MW·m$^{-2}$K$^{-1}$ for the [FePt/h-BN(2.5nm)/FePt] interface at room temperature. We attribute the ultralow thermal conductivity to the weak van der Waals bonding at h-BN/FePt interfaces, which dominates the thermal resistance of the multilayer structure. These findings provide insights into the thermal transport in 2D-material/metal multilayer nanostructures and suggest the [h-BN/FePt] superlattice as a promising material for nanoscale thermal barrier coating. Furthermore, the obtained TBC lays the foundation for analyzing heat transfer in FePt-(h-BN) nanogranular films, a promising magnetic recording media which can potentially provide high thermal gradient for heat-assisted magnetic recording (HAMR). This work advances the understanding of thermal transport in 2D-material/metal nanocomposites and demonstrates interface engineering as an effective approach to achieve materials with ultralow thermal conductivity.

**Topics:** nanocomposite, superlattice structure, thermal conductivity, thermal boundary conductance, multilayer thin film

## I. INTRODUCTION

Layered two-dimensional materials (2DM) exhibit highly anisotropic thermal conductivity due to their unique bonding mode. The strong intralayer covalent or ionic bonds render high phonon conductivity within each monolayer, while the weak interlayer van der Waals interactions suppress the phonon transmission across the planes. Bulk graphite is an anisotropic heat conductor with a thermal conductivity along the basal planes of $\kappa_{\parallel} \approx 2000$ W·m$^{-1}$K$^{-1}$, whereas its cross-plane thermal conductivity is two orders of magnitude smaller with $\kappa_{\perp} \approx 20$ W·m$^{-1}$K$^{-1}$, both at room temperature.[1] The few-layer h-BN nanosheet exhibits an even higher thermal conductivity anisotropy with $\kappa_{\parallel} \approx 360$ W·m$^{-1}$K$^{-1}$ and $\kappa_{\perp} \approx 2$ W·m$^{-1}$K$^{-1}$, giving rise to $\kappa_{\parallel}/\kappa_{\perp}$ as high as ~180.[2,3] As h-BN is an insulator, phonons dominate the heat transfer in it. The weak interlayer van der Waals forces limit high-frequency vibration modes, which significantly suppresses the phonon transmissivity along the cross-plane directions.[4–6]

Moreover, heat transfer through interfaces comprising 2DM has also attracted great attention, since 2DM have been widely investigated for potential applications in electronic, photonic or spintronic devices. For example, the few-layer h-BN has been used as the dielectric layer in resistive random-access memory (RRAM)[7], and as the tunnel barrier layer in magnetic tunnel junction (MTJ)[8]. Thermal management for these device applications has driven research on the thermal boundary conductance(TBC) of contacts formed by 2DM and other interfacing materials, such as metal or semiconductor. It has been demonstrated that TBC is closely related to the bonding between two materials at the interface[9,10]. In these cases, the nature of interaction between the electrode material and 2DM could determine the thermal conductivity across the interface.

An important experimental model for studying the interfacial thermal transport is multilayer thin films[10]. Multilayers, also known as superlattices or nanolaminates, are structures composed of alternating layers of (most commonly) two materials in a periodic sequence (i.e., two layers per period, ABAB stacking). They have inspired extensive study related to ultralow cross-plane thermal conductivities[11]. Besides their relevance to fundamental study of phonon transport, multilayer structures have found various applications, such as thermal barrier coating[12], soft X-ray mirrors[13] and thermoelectric devices[14–16]. Despite the growing interest in 2DM and their unique thermal properties, the study of multilayers composed of metals and 2DM remains relatively unexplored. By examining the multilayer thin film as a model, one can determine the TBC of interface between the two alternating materials. This is critical for understanding the thermal behavior of more complicated nanocomposite structures, such as nanogranular thin films with metal nanoparticles embedded in other matrix materials. These nanogranular thin films also find widespread applications, such as magnetic recording media[17,18], catalyst[19], and floating gate memory[20].

Recently, the FePt-(h-BN) core-shell nanogranular thin film has been fabricated via sputtering as a promising recording media layer for heat assisted magnetic recording (HAMR) in hard disk drive applications. The film comprises a single layer of $L1_0$ chemically ordered FePt grains and h-BN nanosheets in the grain boundary regions[21–23], as shown in Fig. 1(d) and 1(e). Each FePt grain is essentially encircled by multiple h-BN monolayers which are stacked in the radial direction and conform to the side surfaces of the grain. The in-plane thermal transport between two adjacent FePt grains is, thus, governed by the heat conduction across the h-BN basal planes. The low perpendicular thermal conductivity through the h-BN monolayers($\kappa_\perp$) should provide excellent thermal insulation between adjacent FePt grains, a key feature for achieving high thermal gradient during recording to achieve high recording densities[24–27]. Noting that the interfacial thermal conductance across the FePt grain surface and the encircling h-BN layers is critical, it is referred to as the TBC of FePt/h-BN interface ($h_I$).

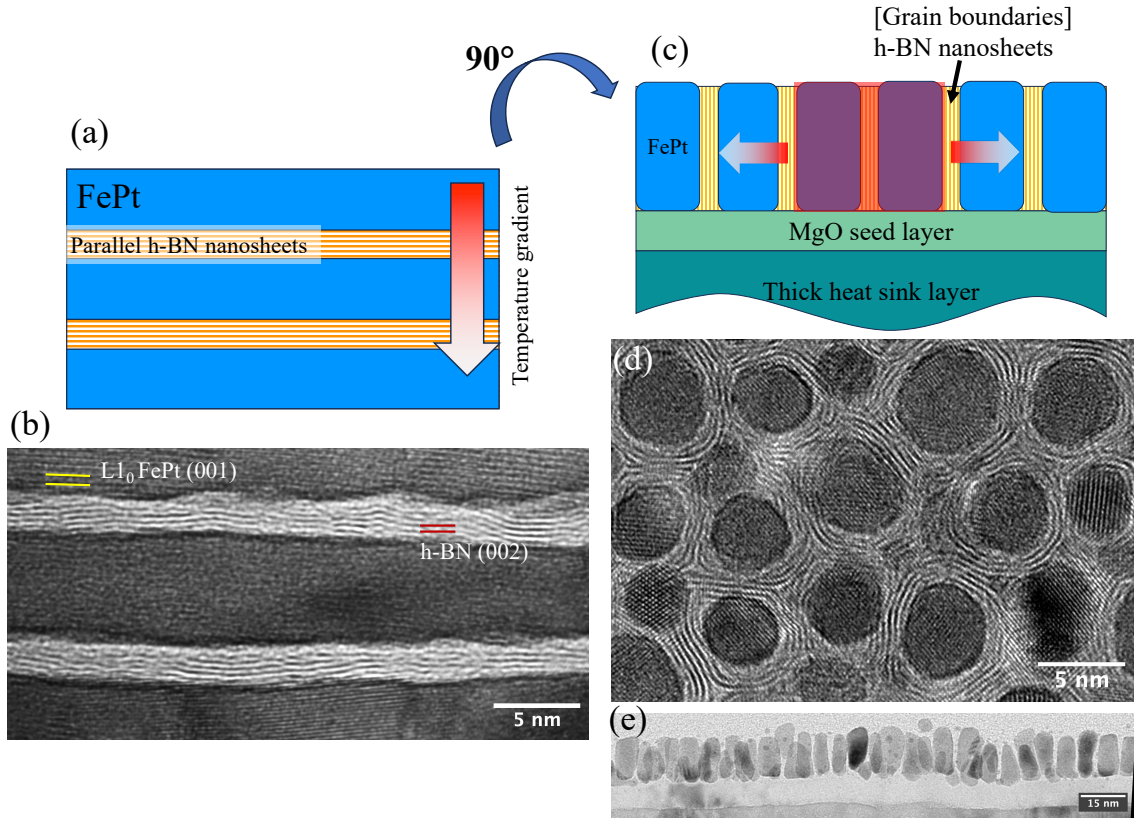

**Figure 1.** (a) schematic structure of the [h-BN/ $L1_0$-FePt] multilayer nanofilm; (b) cross-sectional TEM image of $L1_0$-FePt layers separated by parallel h-BN(002) nanosheets; (c) schematic of heat conduction in FePt-(h-BN) nanogranular film; (d) plane-view HRTEM image and (e) cross-sectional TEM image of the FePt-(h-BN) granular media

In this work, we employed the [h-BN/ $L1_0$-FePt] multilayer structure as a model to study the thermal conductivity of this material system, including the TBC of few-layer h-BN interfacing with FePt. The effective thermal conductivity of the [h-BN/ FePt]×$N$ multilayer was measured using the three-omega (3ω) method. The findings from this work can be extrapolated to analyze the nanogranular thin film of FePt-(h-BN), since it can be considered as a similar system rotated by 90° as illustrated in Fig. 1. Heat transfer within the film plane of the granular FePt-(h-BN) nanostructure passes through high-density interfaces between FePt and bent h-BN (002) basal planes, as depicted in Fig. 1(c) and 1(d). Similar conditions apply to the cross-plane heat flux in the multilayer counterpart, [h-BN/ FePt]×$N$, with an interface density of 0.22 nm$^{-1}$.

## II. EXPERIMENT METHODS

In this study, all multilayer samples were deposited on 4-inch Si (001) substrates using a magnetron sputtering system (AJA international) with base pressures of 2×10$^{-8}$ Torr. Each sample possesses an adhesion layer of Ta (2 nm), followed by an underlayer of Ru (8 nm) to ensure flat surfaces during the repeated heating for h-BN deposition. The thin film stacks that underwent measurements were [h-BN (2.5±0.2 nm)/ L1$_0$-FePt (6.5±0.2 nm)]×$N$ alternating multilayer thin films, where $N$ = 4, 7, 10, 13, representing the number of periods of [h-BN/FePt] bilayers in the film stacks. The samples are named as 4L, 7L, 10L and 13L, respectively. Deposition rates were determined by cross-sectional transmission electron microscope (TEM) images of calibration thin film samples. The sputtering conditions are summarized in Table I.

**Table I.** Sputtering Conditions for [h-BN/ L1$_0$-FePt] Multilayer

| Materials | Pressure (mTorr) | Target-substrate distance (cm) | Temperature (°C) | Power (W) | Rate (nm/s) |
|---|---|---|---|---|---|
| FePt | 6 | 10 | Room Temperature | 20 (DC) | 0.015 |
| h-BN | 3 | 4 | 700 | 150 (RF) | 0.010 |

For the formation of h-BN nanosheets at 700°C, an RF substrate bias of power 3W was applied to all BN deposition steps, with bias voltages typically ranging from -4 to -7V. Furthermore, since FePt layers deposited on h-BN surfaces at elevated temperature tend to form islands, the substrates were cooled after each h-BN deposition to near room temperature before the deposition of FePt to achieve continuous and flat FePt layers. A passivation layer of SiO$_2$ (100 nm) was sputtered over the top of each multilayer at room temperature to prevent surface oxidization and minimize shunting of the excitation AC current. The entire film stack is shown schematically in Fig. 3(b). The superlattice structures and crystallography of the multilayer nanofilms were characterized using cross-sectional TEM imaging (FEI Tecnai F20 operated at 200 kV) and X-ray reflectivity/diffraction (XRR/XRD) measurements with Cu Kα radiation (Malvern Panalytical Empyrean diffractometer with mirror optics). The cross-sectional TEM samples were prepared through mechanical polishing, dimpling and ion-milling procedures.

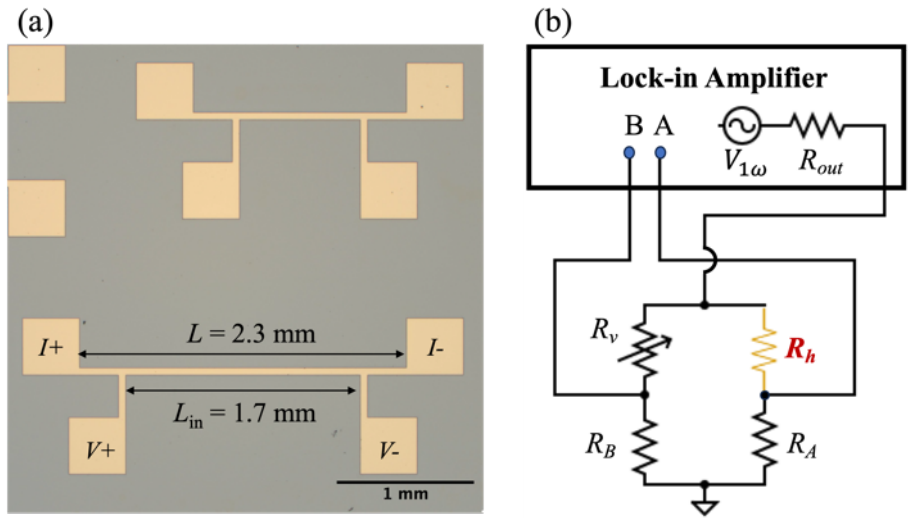

**Figure 2.** (a) Top-view optical micrograph of the Au heater/thermometer devices; (b) diagram of the measured Wheatstone bridge circuit.

Thermal conductivities perpendicular to the film plane were measured using the frequency domain differential three-omega(3ω) method. Figure 2(a) shows the patterned heater filament and Figure 2(b) presents the schematics of the measurement setup. The heater layer was a Ti(2nm)/Au(150nm) bilayer deposited using E-beam evaporation on to the lithographically patterned photoresist and was patterned using lift-off technique. The Au heater filament has a length ($L$) of 2.3 mm and width ($W$) of 50 μm. It was verified by XRR measurements after the heater patterning procedures that the superlattice structure was not affected. It is important to note that the width of the heater (50 μm) is much larger than the total thickness of measured films (150~200 nm), which enables the use of 1D heat conduction approximation in the differential 3ω analysis, ignoring the lateral heat loss in the film plane[28].

Specifically, measurements were conducted not only on the test samples (substrate + underlayer + [h-BN/FePt]×$N$ + passivation), but also on a reference sample (substrate + underlayer + passivation). It is noteworthy that the underlayer of the reference sample, Ta(2 nm)/Ru(8 nm), was also annealed at 700°C for 1 hour and completely cooled down before depositing Ta(2 nm)/SiO$_2$(100 nm) capping layers to be more similar to underlayers of the test samples. The difference in temperature responses of heater between the test sample and the reference sample is attributed to the thermal resistance of the [h-BN/FePt]×$N$ multilayer. The SR830 lock-in amplifier was used to measure the third harmonic signal generated by the 3ω heater. Moreover, to eliminate spurious 3ω noise from the built-in reference voltage source of the amplifier, a Wheatstone bridge circuit was employed, as depicted in the circuit diagram Fig. 2(b). $R_A$ and $R_B$ are two fixed resistors, $R_h$ represents the 3ω heater (wires were connected to two outer contact pads of the four-contact-pad structure), and $R_v$ is a potentiometer which can be adjusted to balance the Wheatstone bridge.

## III. RESULTS AND DISCUSSION

## A. Microstructure of [h-BN/FePt] Multilayer

The well-formed superlattice structure with smooth and continuous interfaces between $L1_0$-FePt and h-BN (002) layers across the entire multilayer sample is evident from both the XRR patterns and the cross-sectional TEM images. Figure 3(a) shows the bright-field (BF) cross-sectional TEM image of the fabricated [h-BN/FePt] multilayer sample (13L), with a periodic thickness of about 9.3 nm. The h-BN and FePt layers are well separated with sharp interfaces. The initial h-BN layer is flat with very smooth FePt/h-BN interfaces. However, as the film grows thicker and the number of repeats increases, the roughness of the h-BN/FePt interfaces increases gradually. Though the interface becomes rough, the h-BN layers still appear conformal to the FePt interfaces. Both $L1_0$ FePt and h-BN layers are polycrystalline. The average grain size of FePt is measured to be around 20nm in the cross-sectional TEM image Fig. 1(b). By comparison, the size of the single-crystal h-BN flakes grown on the FePt layer is much larger, typically ~500 nm measured via exfoliation and TEM. More importantly, Fig. 1(b) clearly shows that the biased RF sputtering of h-BN has developed a (002) texture within each layer. Within each 2.5-nm-thick h-BN (002) layer, there are 5-7 parallel h-BN nanosheets with interlayer distances ranging from 0.33 to 0.35 nm. The variation in interlayer spacing can result from point defects, stress or misaligned stacking of h-BN monolayers[29].

Having h-BN nanosheets parallel to the metal surface is crucial, since this ensures that van der Waals forces dominate the bonding at the interfaces, which likely contributes to a lower TBC. The above observations are consistent with our previous study which has demonstrated that sputtered h-BN thinner than 3nm adopts the layer-by-layer growth mode[22]. In this study, the thickness of the h-BN layer in each period is kept the same, $d_{h-BN} = 2.5$ nm. This is because as the thickness of the sputtered h-BN film increases, the h-BN nanosheets gradually turn to grow perpendicular to the film plane.

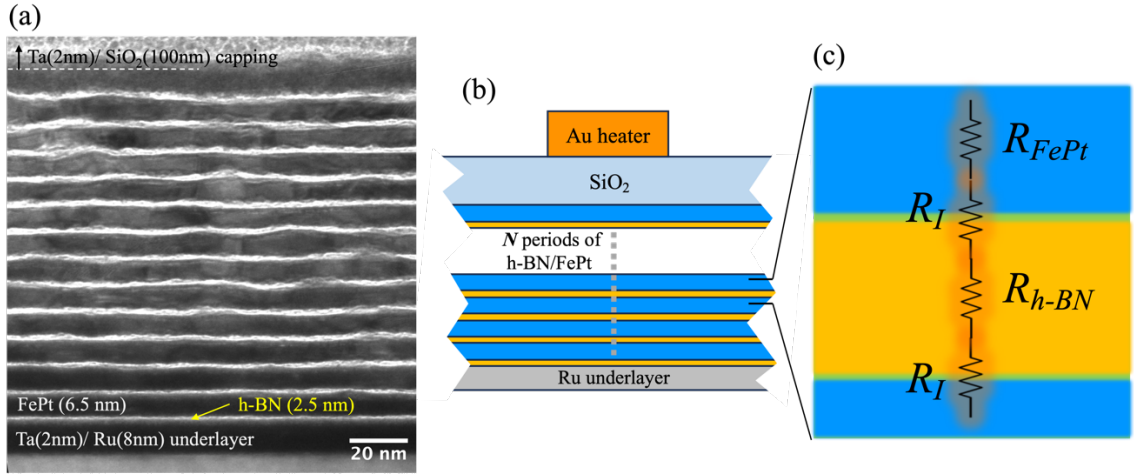

**Figure 3.** (a) cross-sectional TEM image of multilayer sample 13L; (b) schematic of the entire film stack; (c) thermal circuit diagram of a period in the [h-BN/FePt] multilayer structure.

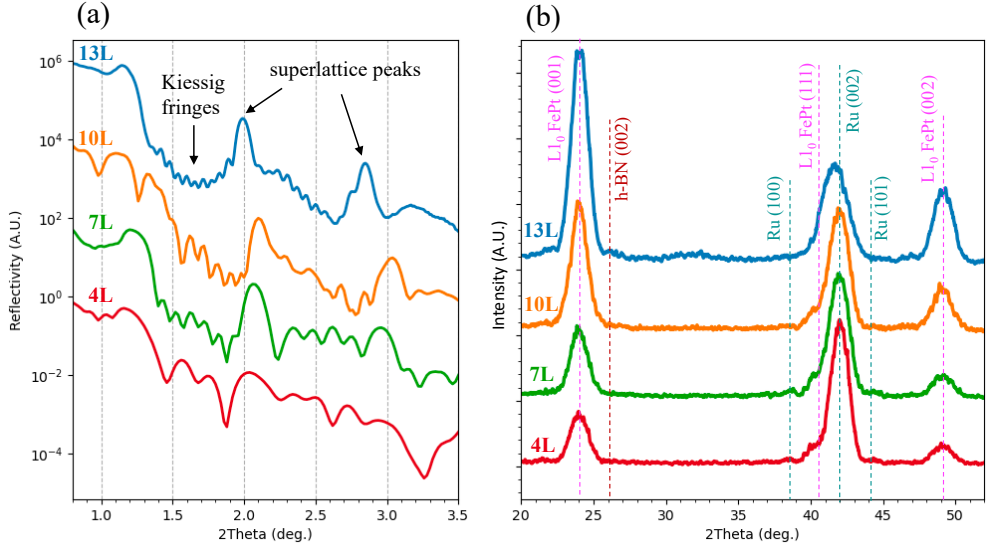

**Figure 4.** (a) XRR patterns and (b) XRD patterns of all [h-BN/FePt]×$N$ multilayer nanofilms

The XRR patterns of the multilayer samples, Fig. 4(a), show clear signs of the presence of superlattice structures. For alternate stacks of nanoscale layers of two materials, the interference of X-rays reflected from different layers generates a distinct pattern featuring relatively intense peaks (Bragg-like superlattice peaks) separated by a few smaller peaks (Kiessig fringes)[16]. The number of Kiessig peaks is directly related to the number of repetition periods (*N*), ideally equals to *N*-2. From Fig. 4(a), the superlattice peaks are clearly observed in sample 7L, 10L, and 13L, while they are less well-defined in the pattern of sample 4L. The counts of Kiessig peaks are 2, 5, 8 and 10 for sample 4L, 7L, 10L and 13L, respectively. The absence of one Kiessig peak in the thickest sample 13L may be attributed to cumulative roughness, which degraded the interface formed in the last [h-BN/FePt] period on the top. The period thickness in the range of 8.7 ~ 9.5 nm calculated from the curve fittings agrees well with the cross-sectional TEM images. Overall, the XRR patterns confirm the presence of superlattice nanostructures with sharp and continuous [h-BN (002)/ $L1_0$-FePt] interfaces in larger areas of the multilayer samples.

The out-of-plane XRD spectra of the [h-BN/FePt] multilayer samples are displayed in Fig. 4(b). The hexagonal close-packed (hcp) Ru underlayers have developed a good (002) texture on the amorphous Ta(2 nm) layer in all samples. More importantly, intense $L1_0$-FePt (001) and (002) peaks, along with relatively small FePt (111) peaks, are observed in all samples, whose intensities scale directly with the number of periods *N*. The large intensity ratio of the $L1_0$-FePt (001) superlattice peak to the (002) fundamental peak indicates highly ordered $L1_0$-FePt formed in the layers. The majority of FePt grains are (002) textured, with a relatively small portion of (111) textured grains. The FePt (111) peaks appear as shoulder peaks next to the Ru (002) peaks in samples 4L, 7L, 10L, and eventually merge with the Ru (002) peak in sample 13L, forming a broader peak in between the peak positions of the two component peaks. Notably, two additional small humps are present in sample 13L centered at 26° and 32°, corresponding to the d-spacings of h-BN(002) and $L1_2$ FePt$_3$(110) planes respectively. The Pt-rich $L1_2$-FePt$_3$ phase was also observed in a previous study of FePt co-sputtered with BN at high temperature[30].

**B. Cross-plane Thermal Conductivity and TBC**

The third harmonic signal (in-phase component) in the Wheatstone bridge output detected by the lock-in amplifier, $V_{A-B,3\omega}$ can be converted to root-mean-square(rms) $3\omega$ voltage oscillation across the Au heater by: $V_{h,3\omega} = \frac{R_A+R_{h,0}}{R_A} \times V_{A-B,3\omega}$, where $R_A$ is the fixed resistor and $R_{h,0}$ is the room-temperature resistance of the three-omega heater in branch A of the Wheatstone bridge, as depicted in the electric circuit diagram Fig. 2(b). The driving voltage drops across the heater, $V_{h,1\omega}$, were also measured. The $2\omega$ temperature oscillation (rms amplitude) on the three-omega heater can be calculated as $\Delta T_{h,2\omega} = \frac{2}{\alpha} \times \frac{V_{h,3\omega}}{V_{h,1\omega}}$, where $\alpha$ represents the thermal coefficient of resistance(TCR) of the Au heater. The TCR of Au heater was pre-calibrated using 4-probe measurements in a temperature range of 290K to 315K, giving a value of $(3.21 \pm 0.02) \times 10^{-3}$ K$^{-1}$. As shown in Fig. 5(a), the plots of $\Delta T_{h,2\omega}$ versus driving frequency (log-scale) over the range of 100~ 2000 Hz exhibit highly linear trends for all samples. All data lines of the test samples are parallel to that of the reference sample because the slope reflects the thermal conductivity of Si(001) substrates, which is known as the three-omega slope method. The thermal conductivity of Si substrates is calculated to values ranging from 145 to 152 W·m$^{-1}$K$^{-1}$, consistent with the reported value of 147 W·m$^{-1}$K$^{-1}$ for Si [28]. This confirms the reliability of our three-omega measurement setups.

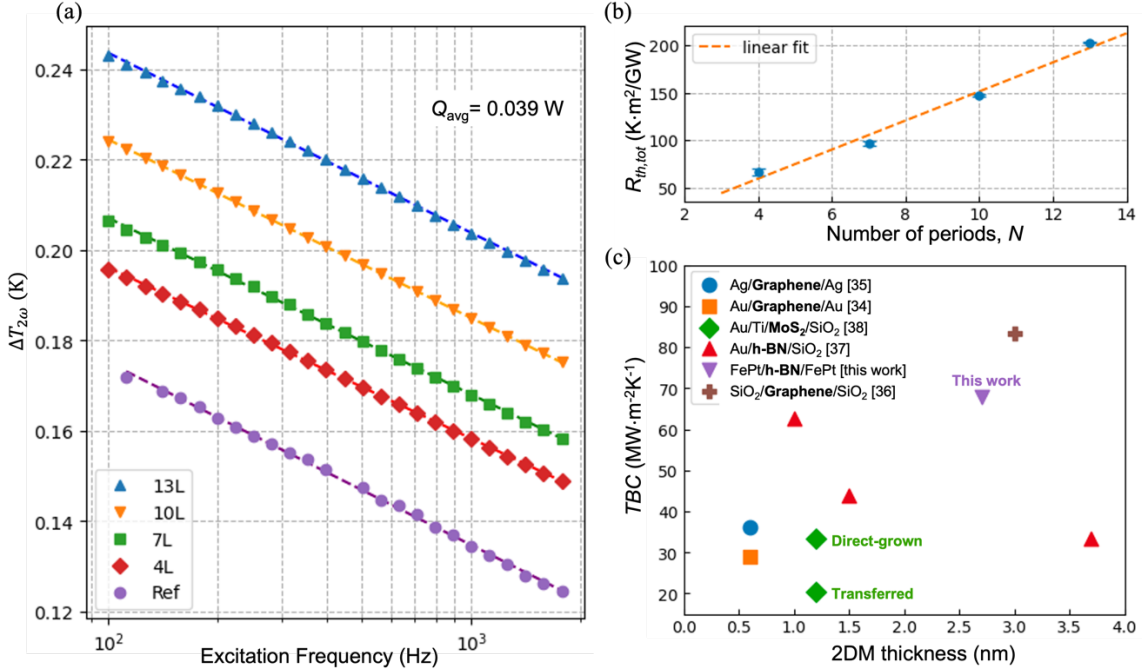

**Figure 5.** (a) Semi-log plot of temperature oscillation in the three-omega thermometer as a function of the driving current frequency (log scale); (b) total thermal resistance of [h-BN/FePt]×$N$ multilayers (calculated from differential three-omega measurements) as a function of number of period $N$; (c) Summary of TBC values for selected [X/2DM/Y] contacts with the corresponding thickness of 2D materials.

As described in the experimental section, the cross-plane thermal conductivities of multilayer samples were analyzed using the differential three-omega method. The offset values of heater-temperature oscillations between the test samples and the reference sample, $\Delta T_{h,test} - \Delta T_{h,ref}$, are independent of the driving frequency (1$\omega$) in this linear regime. The total thermal resistances of the [h-BN/FePt]×$N$ multilayers were calculated as

$$R_{th,tot} = A \times \frac{\Delta T_{h,test} - \Delta T_{h,ref}}{Q_{\text{avg}}}, \qquad (1)$$

where $Q_{avg} = \frac{V_{h,1\omega}^2}{R_{h,0}}$ represents the average heating power, and $A = WL$ represents the area covered by the Au filament. The total thermal resistance is directly related to the number of periods $N$ as plotted in Fig. 5(b). The slope of the linear fitting should be

$$\frac{\delta R_{th,tot}}{\delta N} = R_{FePt} + R_{h-BN} + 2R_I, \quad (2)$$

since each repeating unit, the [h-BN/FePt] bilayer, contributes this much into the total thermal resistance, as shown in the thermal circuit diagram in Fig. 3(c). The terms on the right side of the above equation are thermal resistances for each of the FePt layer, h-BN layer, and the FePt/h-BN interface, respectively. In specific,

$$R_{FePt} = \frac{d_{FePt}}{\kappa_{L1_0\ FePt}}, \quad (3)$$

$$R_{h-BN} = \frac{d_{h-BN}}{\kappa_{h-BN\perp}}, \quad (4)$$

$$R_I = \frac{1}{h_I}, \quad (5)$$

where $d_{FePt}$ and $d_{h-BN}$ are thicknesses of FePt and h-BN in a single period, $\kappa_{L1_0\ FePt}$ and $\kappa_{h-BN\perp}$ represents the thermal conductivity of L1$_0$-ordered FePt and few-layer h-BN (002) along c-axis, respectively, $h_I$ represents the TBC of the [h-BN (002) /L1$_0$-FePt] interfaces. This slope method avoids the unknown TBC values for Ru/h-BN interface, FePt/Ta/SiO$_2$ interface and Ru/Ta/SiO$_2$ interface, which jointly determine the intercept of data lines on y-axis.

The reported values of $\kappa_{L1_0\ FePt}$ is 11.5±0.8 W·m$^{-1}$K$^{-1}$,[31] and $\kappa_{hBN\perp}$ is 2.0±0.2 W·m$^{-1}$K$^{-1}$ [32] at room temperature. Therefore, the TBC of the [h-BN (002) /L1$_0$-FePt] interface, $h_I$, is estimated as 149.9±16.0 MW·m$^{-2}$K$^{-1}$ at room temperature. This value is comparable to that of the Al/Al$_2$O$_3$ interface[33]. It is worth noting that the interfacial thermal resistance ($2NR_I$) accounts for more than 80% of the total resistance $R_{th,tot}$ in all multilayer samples. This indicates phonon scattering at the [h-BN (002) /L1$_0$-FePt] interfaces dominates the heat resistance across the multilayer nanofilms. Some reports considered the [X/2DM/Y] encapsulated structure as a quasi-2D interface, where X and Y represents two materials that sandwich the few-layer/monolayer 2D materials. To compare with the existed results, the TBC of [FePt/h-BN(2.5 nm)/FePt] structure can be defined, $h_{FBF}$ = 67.9 ± 6.6 MW·m$^{-2}$K$^{-1}$.

Low TBC value is a characteristic of the low-stiffness, van der Waals-bonded interfaces formed between 2D materials and 3D bulk materials. A summary of TBC values for various [X/2DM/Y] contacts reported to date, with the corresponding thickness of 2D materials, is presented in Fig. 5(c). Many studies of [Metal/monolayer-Graphene/Metal] interface[34,35] show TBC values ranging from 20 to 40 MW·m$^{-2}$K$^{-1}$. Z. Chen, et al. measured the [SiO$_2$/Graphene(3nm)/SiO$_2$] interface[36], giving 83 MW·m$^{-2}$K$^{-1}$. Xinxia Li, et al. reported measured TBC of [Au/h-BN(3.7nm)/SiO$_2$] using 3$\omega$ method with a value of about 33.3 MW·m$^{-2}$K$^{-1}$ at 300K[37]. The $h_{FBF}$ value reported here is relatively high, potentially because h-BN nanosheets are direct-grown on FePt, instead of being exfoliated and transferred. Direct-grown 2DM normally exhibit stronger binding energies with substrate than transferred ones which are merely physisorbed on the surface. Poya Yasaei et al. also reported a markedly higher TBC for the [Au/Ti/MoS$_2$/SiO$_2$] interface with direct-grown MoS$_2$ (33.5 MW·m$^{-2}$K$^{-1}$) than the counterpart with transferred MoS$_2$ layer (20.3 MW·m$^{-2}$K$^{-1}$)[38]. Additionally, this work offers a good estimate of the effective TBC assuming particle behavior of phonon transport. However, with a period thickness of only 9 nm in the multilayer structure, coherent phonon transport may become significant, which would lead to an overestimated TBC.

The effective thermal conductivity of the entire [h-BN/FePt]×$N$ multilayer, ignoring the superlattice nanostructures, was estimated as

$$\kappa_{ML} = \frac{d_{tot}}{R_{th,tot}}, \tag{7}$$

where $d_{tot}$ is thickness of the whole [h-BN/FePt]×$N$ multilayer (Ru underlayer excluded). The results of $\kappa_{ML}$ values obtained from the four multilayer samples show good consistency, giving an average value of 0.60±0.05 W·m$^{-1}$K$^{-1}$. At room temperature, this $\kappa_{ML}$ is comparable to the reported thermal conductivities of Au/Si[11], W/Al$_2$O$_3$[12] and Mo/Si[13] multilayers with similar interface densities ranging from 0.20 to 0.25 nm$^{-1}$. Unlike Mo/Si and Au/Si multilayers, where atomic interdiffusion degrades the superlattice structures, the [h-BN/FePt]×$N$ multilayer exhibits exceptional thermal stability. Therefore, this [h-BN/FePt]×$N$ multilayer holds potential as a nanometer-scale thermal barrier coating.

This quantitative analysis for thermal characteristics in the h-BN/FePt system obtained here provides important understanding for evaluating the in-plane thermal conductivity of L1$_0$ FePt-(h-BN) granular films as HAMR media. In HAMR, FePt grains are heated by extremely localized electric field at optical frequency, and it is critical to have the heat stay local. Lateral thermal conductivity, more precisely the thermal conductivity between adjacent FePt grains, needs to be sufficiently small. The plane-view TEM image in Fig. 1(d) and the cross-sectional one in Fig. 1(e) illustrate the microstructure of FePt-(h-BN) films, where the column-shaped L1$_0$-ordered FePt grains (diameter < 7nm) are fully encircled by h-BN nanosheets (width ~1nm). The side surface of a FePt grain and the surrounding h-BN layers form a circumferential [h-BN(002)/FePt] interface, and adjacent FePt grains are all separated by the h-BN nanosheets. The TBC values obtained in this work can, therefore, be directly applied to estimate the thermal conductivity between adjacent FePt magnetic grains. The small radius of the FePt grains, ~ 3 nm, is likely to yield even smaller thermal conductance in comparison to the one obtained with h-BN layers in between two large parallel FePt planes ($h_{FBF}$).

Nanogranular FePt thin films can be formed with other grain boundary materials, such as amorphous carbon (a-C)[39] and silicon oxides(SiO$_x$)[40]. Hoan Ho et al. measured a similar multilayer of [FePt (14.5 nm)/ a-C (1 nm)]×*N* using the frequency-domain thermoreflectance (FDTR) method[26], where the thermal resistance through a 1-nm-thick carbon layer and the two a-C/FePt interfaces was reported as $R_{C(1nm)} + 2R_{I,CF} \approx 6.7 \text{ K} \cdot \text{m}^2/\text{GW}$. In comparison, the thermal resistance through a l-nm-thick h-BN and the two h-BN/FePt interfaces can be calculated: $R_{hBN(1nm)} + 2R_{I,BNF} \approx 13.8 \text{ K} \cdot \text{m}^2/\text{GW}$. In other words, the FePt-(h-BN) granular media should be at least twice as more thermally resistive compared to the FePt-C granular media in terms of lateral thermal conduction between adjacent grains.

**IV. CONCLUSION**

In summary, we have demonstrated a method to fabricate a multilayer nanofilm comprising alternate stacking of FePt (6.5 nm) and h-BN(002) nanosheets (2.5 nm). TEM imaging and analysis of XRD, XRR patterns revealed a well-defined superlattice nanostructure with sharp interfaces, showcasing highly L1$_0$-ordered FePt layers separated by few-layer h-BN nanosheets parallel to the metal surfaces. In order to sputter h-BN nanosheets in a layer-by-layer growth mode on metal surfaces, moderate substrate bias voltage (-5~-15V) and high deposition temperature (> 500°C) must be employed. Thermal measurements using three-omega method demonstrated a linear relationship between multilayer heat resistance and number of periods, whose slope provides an estimated TBC of the [L1$_0$-FePt/h-BN/L1$_0$-FePt] interface, $h_{FBF}$ = 67.9 ± 6.6 MW·m$^{-2}$K$^{-1}$. The interface thermal resistance dominates the total cross-plane resistance in this superlattice nanostructure. The entire multilayer nanofilm of [h-BN/ FePt]×*N* exhibits an ultralow value of effective thermal conductivity, 0.60±0.05 W·m$^{-1}$K$^{-1}$, with an interface density of about 0.22 nm$^{-1}$.

The weak interfacial bonding dominated by van der Waals force is considered to be the main contributor to the ultralow thermal conductance, which can lead to strong phonon scatterings at the interfaces. The [h-BN/ FePt]×*N* multilayer shows promise for applications such as nanoscale thermal barrier coating with good thermal stability. Moreover, this investigation of the model multilayer system can provide fundamentals for studying thermal conduction in another particulate composite structure, the FePt-(h-BN) nanogranular thin films.

**ACKNOWLEDGMENTS**

This research was funded by the Data Storage Systems Center at Carnegie Mellon University and all its industrial sponsors and by the Kavcic–Moura Fund at Carnegie Mellon University. The authors would like to thank Xinyi Fang, Benjamin Friedman, Jinchen Han, Prof. Jonathan Malen and Prof. James A. Bain for helpful discussions. The authors acknowledge the use of the Materials Characterization Facility at Carnegie Mellon University supported by Grant No. MCF-677785.

**AUTHOR DECLARATIONS**

**Conflict of Interest**

The authors have no conflicts to disclose.

**Author Contributions**

**Chengchao Xu**: Conceptualization, Data Curation(lead), Formal Analysis, Methodology (lead), Writing– original draft(lead). **Enbo Zhang**: Data Curation, Methodology, Writing – review and editing. **Bo-Yuan Yang**: Data Curation, Methodology. **B.S.D.Ch.S Varaprasad:** Conceptualization, Methodology. **David. E. Laughlin**: Supervision, Writing – review and editing.

**Jian-Gang (Jimmy) Zhu:** Conceptualization(lead), Methodology, Supervision, Writing – original draft, Funding Acquisition.

## DATA AVALIBILITY

The data that support the findings of this study are available from the corresponding author upon reasonable request.